\def\ctz#1{Ref.~\refcite{#1}}

\def\rfr#1{Eq. (\ref{#1})}

\def\dert#1#2{\frac{{{d}}{#1}}{{{d}}{#2}}}              

\def\bar{\begin{eqnarray}}
\def\ear{\end{eqnarray}}
\def\bb{\bibitem}
\def\eqi{\begin{equation}}
\def\eqf{\end{equation}}
\def\eqia{\begin{eqnarray}}
\def\eqfa{\end{eqnarray}}
\def\rp#1#2{{#1\over#2}}

\def\lb#1{\label{#1}}

\def\apio{A_{\rm Pio}}
\def\vpio{v_{\rm Pio}}
\def\msm#1{ {\rm m}\ {\rm s}^{-{#1}} }

\def\oc2{$\mathcal{O}(c^{-2})$}

\def\bds#1{\boldsymbol{#1}}

\documentclass{ws-ijmpd}

\begin{document}

\title{CAN THE PIONEER ANOMALY BE INDUCED BY VELOCITY-DEPENDENT FORCES? TESTS IN THE OUTER REGIONS OF SOLAR SYSTEM WITH PLANETARY DYNAMICS}

\author{LORENZO IORIO}

\address{INFN-Sezione di Pisa. Viale Unit$\grave{a}$ di Italia 68, 70125 Bari (BA), Italy
\\e-mail: lorenzo.iorio@libero.it}

 \maketitle

\begin{history}
\received{16 June 2008}
\accepted{21 September 2008}
\comby{Jorge Pullin}
\end{history}

\begin{abstract}
In this paper we analyze the impact on the orbital motions of the outer planets of the solar system from Jupiter to Pluto of some velocity-dependent forces recently proposed to phenomenologically explain the Pioneer anomaly,  and compare their predictions (secular variations of the longitude of perihelion $\varpi$ or of the semimajor axis $a$ and the eccentricity $e$) with the latest observational determinations by E.V. Pitjeva with the EPM2006 ephemerides. It turns out that while the predicted centennial shifts of $a$ are so huge that they would have been easily detected for all planets with the exception of Neptune, the predicted anomalous precessions of $\varpi$  are too small, with the exception of Jupiter, so that they are still compatible with the estimated corrections to the standard Newton-Einstein perihelion precessions.  As a consequence, we incline to discard those extra-forces predicting secular variations of $a$ and $e$, also for some other reasons, and to give a chance, at least observationally, to those models predicting still undetectable perihelion precessions. Of course, adequate theoretical foundations for them should be found.
\end{abstract}


\section{Introduction}
The Pioneer anomaly\cite{Nie06} (PA) consists of an unmodelled almost constant and uniform acceleration approximately directed towards the Sun of magnitude
\eqi\apio = (8.74\pm 1.33)\times 10^{-10}\ \msm 2\lb{pioacc}\eqf detected\cite{And98,And02,mark,Ol07} in the radiometric data from the Pioneer 10 (launched in March 1972) and Pioneer 11 (launched in April 1973) spacecraft after they passed the $\approx$ 20 AU threshold moving with speed $v_{\rm Pio} \approx 1.2 \times 10^4\ \msm 1$ along roughly antiparallel escape hyperbolic paths undertaken after their previous encounters with Jupiter ($\approx 5$ AU) and Saturn ($\approx 10$ AU), respectively. Concerning the possibility that PA started to manifest itself at shorter heliocentric distances\cite{Nie05,Nie08}, efforts to retrieve and analyze early data from Pioneer 10/11 are currently being performed\cite{Toth08,PioZARM08}.

The Pioneer spacecraft were particularly well suited for radioscience celestial mechanics experiments because they were spin-stabilized\footnote{This was due to the fact that Pioneer 10/11 were equipped with Radioisotope Thermoelectric Generators (RTG) placed at the end of long
booms to be away from the spacecraft and thereby avoid any radiation damage.}; in practice, they could be regarded as gyroscopes so that  only a few orientation
maneuvers, easily  modeled, were needed every year to keep the antenna pointed towards the Earth.  On the contrary, 3-axis stabilized spacecraft like Voyager 1/2 undergo continuous, semi-autonomous, small gas jet
thrusts to maintain the antenna facing the Earth; as a consequence, their navigation  is not as precise as
that of the Pioneer 10/11.

The attempts performed so far to explain  PA in terms of known effects of gravitational\cite{And02} and/or non-gravitational\cite{Mur99,Kat99} origin were found to be not  satisfactory\cite{And99a,And99b}, so that a vast number of  exotic explanations based on modified models of gravity were proposed (see, e.g., \ctz{And02,Ber04,Dit05,Izz06}, and references therein). If PA is due to some modifications of the known laws of gravity, this must be due to a radial extra-force affecting the orbits of the planets as well, especially those moving in the region in which PA manifested itself in its presently known form. The impact of a Pioneer-like additional acceleration  on the motion of major and minor bodies in the outer regions of the solar system was  recently studied by numerous authors with different approaches\cite{comets,Ior06,Ior07,Tan07,Wal07,Sta08}: it turned out that a constant and uniform extra-acceleration  with the magnitude of \rfr{pioacc} would produce huge secular effects which are neatly absent in the planetary data.

It was recently suggested\cite{Lamm07} that, from a purely phenomenological point of view, test bodies moving in the (outer) solar system could experience velocity-dependent extra-accelerations of the form \eqi A_{v}=-|v_r| \left(\rp{\apio}{v_{\rm Pio}}\right), A_{v}=-v_r \left(\rp{\apio}{v_{\rm Pio}}\right),\lb{extraa1}\eqf and \eqi\ A_{v^2}=-v_r^2 \left(\rp{\apio}{v^2_{\rm Pio}}\right),\ A_{v^2}=-|v_r|v_r \left(\rp{\apio}{v^2_{\rm Pio}}\right)\lb{extraa}\eqf where $v_r$ is the radial component of the test particle's velocity $\bds v$; \rfr{extraa1} and \rfr{extraa} would reduce to \rfr{pioacc} for the Pioneer 10/11 spacecraft whose  velocities  can be assumed entirely radial in the outer regions of the solar system in which PA was detected.  Standish in \ctz{Sta08} put on the test such a hypothesis by fitting huge planetary data sets with the dynamical force models of the latest Jet Propulsion Laboratory (JPL) DE ephemerides modified $ad\ hoc$ according to \rfr{extraa1} and \rfr{extraa}  and examining the results in terms, e.g, of the reliability of the estimated parameters. His conclusion was that the existence of extra-accelerations like those of \rfr{extraa1} and \rfr{extraa} at heliocentric distances $\gtrsim 20$ AU cannot be ruled out by the present-day available data of the outer planets because \rfr{extraa1} and, especially, \rfr{extraa} would induce orbital effects on them too small to be detected. Their existence in the inner regions of solar system is, instead, ruled out.

In this paper we will follow a different approach by using the EPM2006 ephemerides produced by E.V. Pitjeva\cite{Pit08} at the Institute of Applied Astronomy (IAA) of the Russian Academy of Sciences (RAS). First, we will  analytically work out the secular effects  of small perturbing accelerations like those of \rfr{extraa1} and \rfr{extraa} on the Keplerian orbital elements of a planet in order to gain as clear as possible insights about the modifications which the orbits would undergo if \rfr{extraa1} and \rfr{extraa} were real; should some implausible physical feature turn out, it would be more difficult to trust such proposed anomalous forces. Then, we will compare some of such predictions with the latest observational determinations for the outer planets estimated by Pitjeva with the EPM2006 ephemerides in a purely phemomenological way as corrections to the known effects due to usual Newton-Einstein laws, without modelling any additional force.
In Table \ref{Pittable} we quote some quantities we will use. They are the outcome of a global fit of more than 400,000 data points (1913-2006) performed by Pitjeva\cite{Pit07,Pit08} with the EPM2006 ephemerides; about 230 parameters were estimated.
\begin{table}
\tbl{ Second column: formal standard deviations $\delta a$, in m, of the semimajor axes of the outer planets from a fit of 400,000 data points spanning almost one century with the EPM2006 ephemerides (from Table 3 of Ref.~24). Third column: corrections to the standard Newton-Einstein secular precessions of the perihelia$^{25}$ in arcsec cy$^{-1}$. Fourth column:  their formal errors, in arcsec cy$^{-1}$, re-scaled by a factor 10.  For Neptune and Pluto no secular precessions have been estimated because the available modern data records for them do not yet cover an entire orbital revolution.
 }
{\begin{tabular}{@{}cccc@{}} \toprule
Planet & $\delta a$ (m)  & $\Delta\dot\varpi$ (arcsec cy$^{-1}$)& $\delta\Delta\dot\varpi$ (arcsec cy$^{-1}$) \\
\colrule
Jupiter &  615 & 0.0062 &  0.036\\
Saturn & 4,256 & -0.92 & 2.9 \\
Uranus & 40,294 & 0.57 & 13.0\\
Neptune & 463,307 & N.A. & N.A.\\
Pluto & 3,412,734 & N.A. & N.A.\\
\botrule
\end{tabular}\label{Pittable}}
\end{table} %
It must be noted that the uncertainties $\delta\Delta\dot\varpi$ in the estimated corrections to the perihelion precessions are the formal ones re-scaled by a factor 10 in order to obtain realistic evaluations for them.
\section{The orbital effects of velocity-dependent perturbing forces yielding Pioneer-type accelerations}
\subsection{Forces linear in velocity}
According to the classification of \ctz{Sta08}, the first two kinds of extra-forces are linear in $v_r$ being
\eqi A_v^{(2)} = -\left|v_r\right|\mathcal{K},\lb{type2}\eqf
\eqi A_v^{(3)} = -v_r\mathcal{K},\lb{type3}\eqf
with
\eqi \mathcal{K}=\rp{\apio}{\vpio} = 7.3\times 10^{-14}\ {\rm rad}\ {\rm s}^{-1} = 47.4\ {\rm arcsec}\ {\rm cy}^{-1}.\eqf
The radial acceleration of \rfr{type2} is constantly inward, i.e. directed towards the Sun, while the one of \rfr{type3} is directed towards the Sun when $v_r>0$, i.e. when the planets gets farther from the Sun, while is directed away the Sun when $v_r<0$, i.e. when the planet gets closer to the Sun. Indeed, for an unperturbed Keplerian ellipse\cite{Roy}
\eqi v_r = \rp{nae\sin f}{\sqrt{1-e^2}} = \rp{nae\sin E}{1-e\cos E},\eqf where $a$ is the semimajor axis, $e$ is the eccentricity, $n=\sqrt{GM/a^3}$ is the mean motion, $f$ is the true anomaly counted anticlockwise from the perihelion, and $E$ is the eccentric anomaly. $v_r>0$ for $0<f<\pi$, i.e., from the perihelion to the aphelion, and $v_r<0$ for $\pi< f <2\pi$, i.e. from the aphelion back to the perihelion.

In view of the smallness of \rfr{type2} and \rfr{type3} for the planets of the solar system, we will treat them perturbatively. Indeed, the radial velocities for the outer planets amount to $10^1 - 10^3$ m s$^{-1}$ only, so that $A_v\approx 10^{-11}$ m s$^{-2}$, while the Newtonian attraction of the Sun is for them of the order of $10^{-4} - 10^{-6}$ m s$^{-2}$.

Let us work out the secular precession of the longitude of perihelion $\varpi$. The Gauss equation for its variation due to an entirely radial perturbing acceleration $A_r$ is\cite{Roy}
\eqi\dert\varpi t = -\rp{\sqrt{1-e^2}}{nae}A_r\cos f.\lb{gaus}\eqf By inserting \rfr{type2} into \rfr{gaus}, evaluating the r.h.s over the unperturbed Keplerian ellipse characterized by
\begin{equation}
\begin{array}{lll}

r = a(1-e\cos E),\\\\
dt = \left(\rp{1-e\cos E}{n}\right)dE,\\\\
\cos f = \rp{\cos E - e}{1-e\cos E},\\\\
\sin f = \rp{\sqrt{1-e^2}\sin E}{1-e\cos E},
\end{array}\lb{cofi}
 \end{equation}
   and averaging over one orbital period one gets
   \eqi\left\langle\dot\varpi\right\rangle = -\rp{\mathcal{K}\sqrt{1-e^2}}{\pi}\left[\rp{2e - (1-e^2)\ln\left(\rp{1+e}{1-e}\right)}{e^2}\right]< 0.\lb{odot}\eqf We have used the fact that
   \eqi |v_r| = v_r,\ 0\leq f\leq \pi;\ |v_r| = -v_r,\ \pi\leq f \leq 2\pi.\eqf

   Instead, \rfr{type3} yields no perihelion precession. Indeed,
   \eqi \left\langle\dot\varpi\right\rangle = -\rp{\mathcal{K}\sqrt{1-e^2}}{2\pi}\int_0^{2\pi}\rp{(\cos E - e)\sin E}{1-e\cos E}dE = 0.\eqf

The Gauss equation for the variation of the semimajor axis due to the a radial perturbing acceleration is\cite{Roy}
\eqi\dert a t = \rp{2e}{n\sqrt{1-e^2}}A_r\sin f.\eqf
By proceeding as before it turns out that \rfr{type2} does not yield secular variations of $a$; instead, \rfr{type3} induces a secular decrease of $a$ according to
\eqi\left\langle \dot a\right\rangle = 2\mathcal{K}a\left(\rp{1}{\sqrt{1-e^2}}-1\right).\lb{adot}\eqf    Note that for circular orbits, i.e. $v_r=0$, $\left\langle\dot a\right\rangle=0$.
Since
\eqi \dert e t = \left(\rp{1-e^2}{2ae}\right)\dert a t\eqf when $A=A_r$, also the eccentricity decreases:
\eqi \left\langle \dot e\right\rangle = \rp{\mathcal{K}(1-e^2)}{e}\left(\rp{1}{\sqrt{1-e^2}}-1\right).\lb{edot}\eqf
As expected for a central force, the orbital angular momentum $L=\sqrt{GMa(1-e^2)}$ is conserved, on average: indeed,
\rfr{adot} and \rfr{edot} yield
\eqi\left\langle\dert {L^2} t\right\rangle = GM\left[\left\langle \dot a\right\rangle(1-e^2) - 2ae\left\langle\dot e\right\rangle\right] = 0.\eqf
Instead, the energy $\mathcal{E}=-GM/2a$ is not conserved; according to \rfr{adot}, \eqi\left\langle\dot{\mathcal{E}}\right\rangle = \rp{GM}{2a^2}\left\langle\dot a\right\rangle = \rp{GM\mathcal{K}}{a}\left(\rp{1}{\sqrt{1-e^2}}-1\right) <0.\lb{strano}\eqf
Such a result is certainly suspect from a physical point of view.

\begin{figure}
 \centerline{\psfig{file=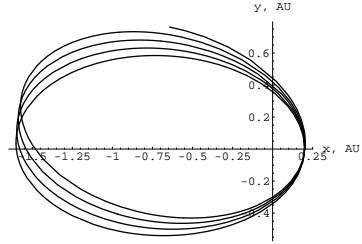,width=4.7cm}}
\vspace*{8pt}
\caption{Numerically integrated trajectory for the radial acceleration of \rfr{type2} proportional to $-|v_r|$. The longitude of perihelion $\varpi$ undergoes a retrograde precession while neither the semimajor axis $a$ nor the eccentricity $e$ experience secular variations.\label{figtype2}}
\end{figure}

\begin{figure}
 \centerline{\psfig{file=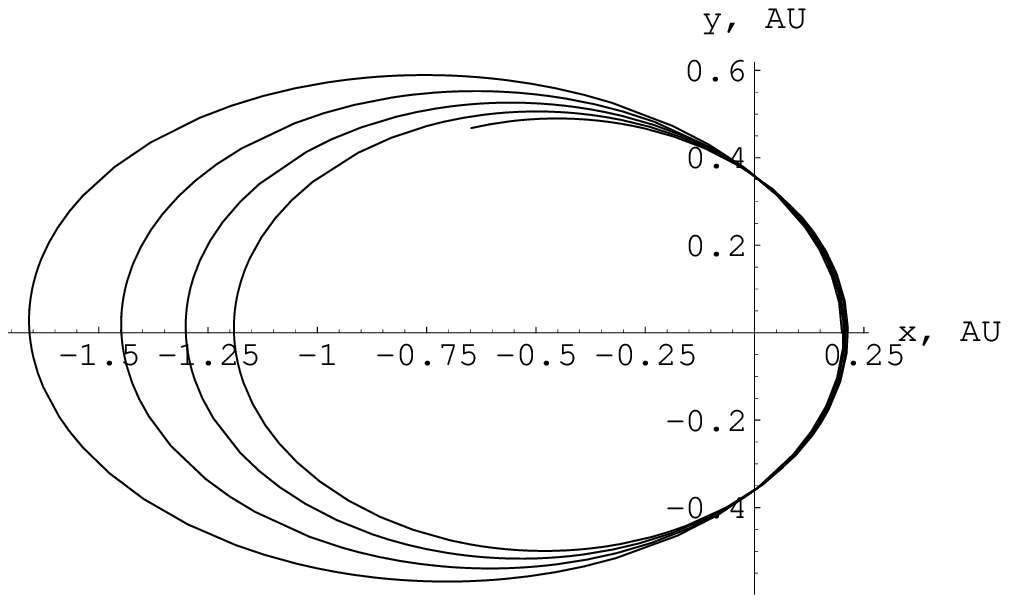,width=4.7cm}}
\vspace*{8pt}
\caption{Numerically integrated trajectory for the radial acceleration of \rfr{type3} proportional to $-v_r$.  Both the semimajor axis $a$ and the eccentricity $e$ secularly decrease, while the longitude of perihelion $\varpi$ remains fixed.\label{figtype3}}
\end{figure}

In order to independently check the results obtained analytically we performed two numerical integrations of the equations of motion adding to the Newtonian monopole term the perturbing accelerations of \rfr{type2} and \rfr{type3}: the qualitative features of the resulting motions are depicted in Figure \ref{figtype2} and Figure \ref{figtype3}.

%
%
%

%
%
%
%
Let us now consider the problem of the existence of the accelerations of \rfr{type2} and \rfr{type3} from a phenomenological point of view according to the present-day planetary data available. In Table \ref{tavola23} we quote the predictions for the outer planets of the centennial shifts in m of the semimajor axis, according to \rfr{adot}, and of the secular perihelion precessions in arcsec cy$^{-1}$, according to \rfr{odot}.
\begin{table}
\tbl{ Second column: shift  $\Delta a$, in m, of the semimajor axis of the outer planets over 1 cy according to \rfr{adot} induced by the acceleration of \rfr{type3} proportional to  $-v_r$. Third column: secular precessions $\dot\varpi$ of the perihelia of the outer planets, in arcsec cy$^{-1}$, according to \rfr{odot} due to the acceleration proportional to $-|v_r|$. Such values are to be compared with those in Table \ref{Pittable}.
 }
{\begin{tabular}{@{}ccc@{}} \toprule
Planet & $\Delta a$ (m)  & $\dot\varpi$ (arcsec cy$^{-1}$)\\
\colrule
Jupiter &  -419,726 & -0.973 \\
Saturn & -1,030,797 &  -1.1 \\
Uranus & -1,470,544 &  -0.9\\
Neptune & -76,220 &  -0.1 \\
Pluto & -88,154,057 & -4.9 \\
\botrule
\end{tabular}\label{tavola23}}
\end{table} %
 Such predictions  must be compared with the observationally determined parameters quoted in Table \ref{Pittable}. Concerning the semimajor axis, the present-day accuracy in determining them would clearly allow to detect shifts as large as those of Table \ref{tavola23} for all planets from Jupiter to Pluto with the exception of Neptune, even by re-scaling the values of $\delta a$ of Table \ref{Pittable} by a factor 10 or more. The situation is less neat for the perihelion precessions. Indeed, it turns out that the present-day accuracy in determining them does not allow to rule out \rfr{odot}, with exception of Jupiter. Thus, we conclude that the acceleration of \rfr{type3} proportional to $-v_r$ is to be considered ruled out by observations; it is true that, in principle, adjusting the ephemerides (without modifying their dynamical force models as done with EPM2006) may absorb the exotic signatures, but we do not believe that could occur because of the huge size of them. Instead, the effects induced by \rfr{type2}, proportional to $-|v_r|$, are still compatible with data.
 Of course, in drawing such conclusions we are tacitly assuming that the Pioneer-type anomalous accelerations of \rfr{type2} and \rfr{type3} exist since one century at least.
 Let us assume that they act since much longer time, say 500 Myr; \rfr{type3} and \rfr{adot}  tell us that, in this case, 500 Myr ago the semimajor axes of the outer planets were equal to 19 AU (Jupiter), 44 AU (Saturn), 68 AU (Uranus), 32 AU (Neptune) and 2,983 AU (Pluto). We have used the simple formula \eqi a_0 = a -\dot a\Delta t,\eqf in which $a_0$ represents the semimajor axis in the past while $a$ denotes its current value.
 Concerning the eccentricities, they would have been larger than 1 according to \rfr{edot} and $e_0 = e -\dot e\Delta t$.
 With regard to the future evolution of the orbits of the outer planets, the time required to circularize their orbits with respect to the present-day values of the eccentricities is  of the order of $8\times 10^5$ yr, provided that the Pioneer-type forces considered here will continuously act upon the planets for a so long time span.
 Of course, issues concerning a theoretical justification for \rfr{type2} and \rfr{type3} remain: suffices it to say that they are not, in general, Lorentz-invariant, as can be straightforwardly shown by using
 \eqi {\bds r}^{'} = \Gamma({\bds r} -{\bds V} t) + (\Gamma -1)\rp{({\bds r}\times  {\bds V})\times \bds V}{V^2},\ t^{'} = \Gamma\left(t - \rp{\bds r\cdot\bds V}{c^2}\right),\eqf
 \eqi{\bds v}^{'} = \rp{1}{1-\rp{\bds v\cdot\bds V}{c^2}}\left[\bds v - \bds V +\left(\rp{\Gamma -1}{\Gamma V^2}\right)\left({\bds v}\times{\bds V}\right)\times  \bds V\right],\eqf
 with $\Gamma = 1/\sqrt{1-V^2/c^2}$.
 The conclusions by Standish\cite{Sta08} are that \rfr{type2} and \rfr{type3} cannot exist for planets up to Jupiter and Saturn, while their existence at heliocentric distances $\gtrsim 20$ AU is virtually undetectable from the motion of Uranus, Neptune and Pluto.

\subsection{Forces quadratic in velocity}
  The other two anomalous radial accelerations examined in \ctz{Sta08}, quadratic in the radial velocity, are
 \eqi A_{v^2}^{(4)} = -v_r^2\mathcal{H},\lb{type4}\eqf
\eqi A_{v^2}^{(5)} = -\left|v_r\right|v_r\mathcal{H},\lb{type5}\eqf
with
\eqi \mathcal{H}=\rp{\apio}{\vpio^2}=6.07\times 10^{-18}\ {\rm m}^{-1}.\eqf
The acceleration of \rfr{type4} is always directed towards the Sun, while the one of \rfr{type5} is inward when the planet moves away from the Sun, while it is directed outwards when the planet approaches the Sun, as in the case of \rfr{type3}.

\begin{figure}
 \centerline{\psfig{file=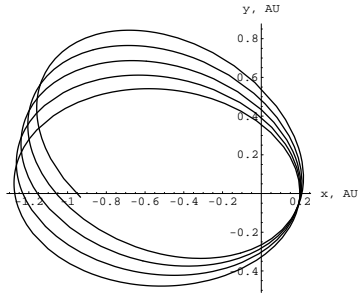,width=4.7cm}}
\vspace*{8pt}
\caption{Numerically integrated trajectory for the radial acceleration of \rfr{type4} proportional to $-v_r^2$.  Both the semimajor axis $a$ and the eccentricity $e$ remain unchanged, while the longitude of perihelion $\varpi$ experiences a retrograde secular precession.\label{figtype4}}
\end{figure}

An acceleration like \rfr{type4} was theoretically obtained by Jaekel and Reynaud in the framework of their linear\cite{JR05a,JR05b} and non-linear\cite{JR06} metric extensions of general relativity. Its orbital effects were worked out in \ctz{Ior07}: neither the semimajor axis nor the eccentricity undergo secular variations, while the longitude of perihelion precesses according to
\eqi\left\langle\dot\varpi\right\rangle = \rp{  {\mathcal{H}} na\sqrt{1-e^2}   }{e^2}\left(-2 + e^2 + 2\sqrt{1-e^2}\right)< 0.\lb{precpio}\eqf
In Figure \ref{figtype4} we show the results of the numerical integration of the equations of motion with \rfr{type4} added to the Newtonian monopole: the results obtained analytically in \ctz{Ior07} are confirmed.

In \ctz{Ior07} it was shown that the  inner planets' perihelion precessions  predicted by \rfr{precpio}  are neatly ruled out by the corrections to the perihelion precessions estimated by Pitjeva\cite{Pit05} with the EPM2004 ephemerides. In Table \ref{neo} we quote the predictions for the outer planets;
\begin{table}
\tbl{ Secular precessions $\dot\varpi$ of the perihelia of the outer planets, in arcsec cy$^{-1}$, according to \rfr{precpio} due to the acceleration of \rfr{type4} proportional to $-v_r^2$. Such values are to be compared with those in Table \ref{Pittable}.
 }
{\begin{tabular}{@{}cc@{}} \toprule
Planet & $\dot\varpi$ (arcsec cy$^{-1}$)\\
\colrule
Jupiter &  -0.030 \\
Saturn &  -0.029\\
Uranus &  -0.015\\
Neptune &  -0.0004\\
Pluto &  -0.308\\
\botrule
\end{tabular}\label{neo}}
\end{table} %
it turns out that they are compatible with the results of Table \ref{Pittable}, apart from Jupiter. Note that it is true also by considering the formal uncertainties in the estimated corrections to the perihelion precessions, i.e. the values  of $\delta\Delta\dot\varpi$ in Table \ref{Pittable} reduced by 10 times. Such a conclusion substantially agrees with that by Standish\cite{Sta08}.

\begin{figure}
 \centerline{\psfig{file=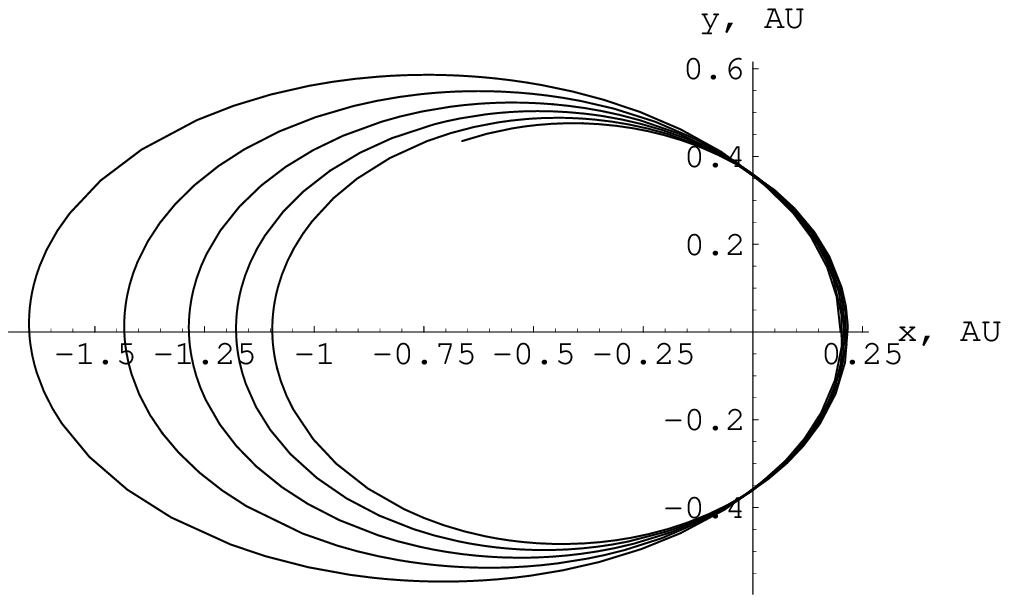,width=4.7cm}}
\vspace*{8pt}
\caption{Numerically integrated trajectory for the radial acceleration of \rfr{type5} proportional to $-|v_r|v_r$.  Both the semimajor axis $a$ and the eccentricity $e$ secularly decrease, while the longitude of perihelion $\varpi$ remains unchanged.\label{figtype5}}
\end{figure}

\rfr{type5}, contrary to \rfr{type4}, induces no secular perihelion precession and a secular variation of the semimajor axis and the eccentricity which decrease according to
\eqi\left\langle\dot a\right\rangle = \rp{4{\mathcal{H}na^2}}{\pi}\left[2e + \ln\left(\rp{1-e}{1+e}\right)\right],\lb{danew}\eqf
\eqi\left\langle\dot e\right\rangle = \rp{2{\mathcal{H}na(1-e^2)}}{\pi e}\left[2e + \ln\left(\rp{1-e}{1+e}\right)\right].\lb{denew}\eqf
Note that $\left\langle\dot a\right\rangle=0$ for circular orbits.  Thus, also \rfr{type5} does not conserve the total energy.
Such analytical results are confirmed by a numerical integration of the equations of motion showed in Figure  \ref{figtype5}.

In Table \ref{tavola5}
we quote the predictions for the centennial semimajor shifts according to \rfr{danew}.
\begin{table}
\tbl{ Shifts of the semimajor axes $a$ of the outer planets, in m, according to the acceleration of \rfr{type5} proportional to $-|v_r|v_r$. Such values are to be compared with those in Table \ref{Pittable}.
 }
{\begin{tabular}{@{}cc@{}} \toprule
Planet & $\Delta a$ (m)\\
\colrule
Jupiter &  -18,753 \\
Saturn &  -39,362 \\
Uranus &  -33,347 \\
Neptune &  -251 \\
Pluto &  -7,283,499 \\
\botrule
\end{tabular}\label{tavola5}}
\end{table}
A comparison with the Table \ref {Pittable} shows that the formal uncertainties $\delta a$ are always quite smaller than such anomalous shifts, apart from Neptune. However, it must taken into account that realistic errors may be up to one order of magnitude larger: if so, it would not be possible to rule out the results of Table \ref{tavola5}, apart from Jupiter. In this case, our conclusions would agree with those by Standish\cite{Sta08}.
Of course, serious issues concerning theoretical justifications of \rfr{type5} and the temporal extent of its existence remain open. Indeed, given the present-day values of the planetary semimajor axes and eccentricities and assuming that \rfr{type5} existed unchanged in the deep past, about 100 Myr-1Gyr ago $e=1$ for the planets from Jupiter to Pluto. Since, instead, the semimajor axes would have remained almost unchanged, this means that the perihelion distances vanished.

\section{Conclusions}
An   ingenious attempt recently proposed to explain the Pioneer anomaly as due to a modification of the usual Newton-Einstein laws of gravitation consists in postulating the existence of some velocity-dependent extra-forces linear or quadratic in the radial component $v_r$ of the velocity of a test body. We put on the test such empirical models in the outer regions of the solar system in which the Pioneer anomaly manifested itself in its presently known form  with the latest observational determinations of the planetary motions obtained by E.V. Pitjeva with the EPM2006 ephemerides. It turns out that the models yielding anomalous perihelion precessions cannot yet be ruled out, at least phenomenologically, for heliocentric distances larger than 5 AU. On the contrary, the models predicting secular variations of the semimajor axis $a$ and the eccentricity $e$ are much more difficult to be trusted not only because they would violate the conservation of energy but also because the centennial shifts for $a$ predicted by them are so large that they should have been  detected, given the present-day accuracy in determining such orbital element. However, it must be considered that sound theoretical justifications for such models must be given.


\end{document}